\documentstyle[aps]{revtex}

\newcommand{\Lag}{{\cal L}}
\newcommand{\DD}{_{{\scriptscriptstyle{1/2}}}}
\newcommand{\LagT}{{\cal L}_{{\rm tr}}}

\newcommand{\ab}[1]{\bar{#1}}
\newcommand{\Four}[1]{\widetilde{#1}}
\newcommand{\prop}{\Delta_F}
\newcommand{\propF}{\Four{\prop}}
\newcommand{\ldp}{\vbox{\ialign{\hfil$##$\hfil\crcr
 \scriptscriptstyle\leftrightarrow\crcr\noalign{\kern.5pt\nointerlineskip}
 \partial\crcr}}}

\begin{document}

\title{Quantum field theory of electrodynamic phenomena that requires no regularization}

\author{Marijan Ribari\v c and Luka \v Su\v ster\v si\v c\cite{email}}

\address{Jo\v zef Stefan Institute, p.p.3000, 1001 Ljubljana, Slovenia}

\maketitle

\begin{abstract}We put forward an example of local, covariant Lagrangians where the Feynman rules result in diagrams of QED but with regularized propagators. Following 't Hooft and Veltman, these diagrams may be taken to define a quantum field theory of electrodynamic phenomena that requires no regularization and is realistic, because: (i)~The corresponding Green's functions are causal. (ii)~Its $S$-matrix is unitary. (iii)~The theory does not imply the existence of particles with wrong metric and/or wrong statistics. (iv)~It contains the experimental predictions of QED. No such Lagrangians were known before.
\end{abstract}
\vskip .5in

\section{Introduction}

When calculating Green's functions and $S$-matrix elements of QED by perturbation expansion, one has to modify (regularize) certain divergent integrals to obtain finite results for Feynman diagrams containing loops. Such regularizations are understood as formalistic technical devices, devoid of any direct physical meaning, that lead upon renormalization to identical, acceptable results. So far no realistic regularization scheme is known that is an integral part of a realistic, physical model of quantum electrodynamic phenomena. Which presents a serious conceptual problem we will propose a solution to\cite{Cao}.

Quantitative predictions about quantum-electrodynamic phenomena implied by the Lagrangian of QED are computed using the Feynman rules, a regularization scheme, and renormalization. Were there a local and covariant Lagrangian where the Feynman rules result in the diagrams of QED but with regularized propagators, we could follow 't Hooft and Veltman\cite{Hooft} and use these Feynman diagrams to define a quantum field theory of electrodynamic phenomena that requires no regularization, provided its Green's functions are causal, and the $S$-matrix is unitary and contains the experimental predictions of QED. Now, there are variants of the Pauli-Villars regularization of QED (where in all diagrams one adds to photon and electron propagators suitable sums of four-vector and bispinor propagators, 't Hooft and Veltman\cite{Hooft} call unitary regulators) that can be derived from such Lagrangians. However, these Lagrangians are not realistic since they imply the existence of heavy particles with wrong metric and/or wrong statistics. It is the purpose of this paper to present for QED an example of realistic, local and covariant Lagrangians that generalize the method of unitary regulators\cite{Hooft} so that no unphysical particles are implied. As far as we know, these Lagrangians\cite{mi001} provide, for the first time since the foundation of QED\cite{Cao}, a framework for realistic quantum field theories of electrodynamic phenomena in four-dimensional space-time with such propagators that no regularization is required.

As to each additional singularity of a unitarly regularized Feynman propagator corresponds an additional particle\cite{Hooft}, we will first show how one can covariantly regularize the Feynman propagators so that they do not acquire additional singularities and have the K\"all\'en-Lehman representation used for proving causality and unitarity\cite{Hooft,Veltm}. Regarding metric and other conventions we follow Refs.~\onlinecite{Hooft,Veltm}; in particular, a four-vector $k = (\vec{k}, ik^0)$, and $k^2 \equiv \vec{k}\cdot \vec{k} - (k^0)^2$.

\section{Covariant regulators without singularities}

Consider a covariantly regularized spin-0 Feynman propagator, say, $\prop(x)$ whose space-time Fourier transform
\begin{equation}
   (2\pi)^4 i\propF(k) = (k^2 + m^2 - i\epsilon)^{-1} 
      + \varphi(k^2 - i\epsilon)
   \label{scpro}
\end{equation}
where: (a)~the regulator $\varphi(z)$ is an analytic function of complex variable $z$ with a finite discontinuity somewhere across the negative real axis; (c)~$|(z+m^2)^{-1} + \varphi(z)| = O(|z|^{-r})$, $r > 1$, as $|z| \to \infty$; (d)~$\varphi(z)$ is real for $z > 0$; (e)~$\varphi(z)$ depends on some cutoff parameter $\Lambda$ so that for any $\Lambda \ge \Lambda_0 > 0$ it has properties (a) to (d), and for $\Lambda \to \infty$,
\begin{equation}
   \sup_{z \ge 0} |\varphi(z)| \to 0
   \label{regpr0}
\end{equation}
and
\begin{equation}
   \sup_{|z| < z_0} |\varphi(z)| \to 0
   \qquad\hbox{for any}\quad z_0 > 0.
   \label{regpr}
\end{equation}
Using Cauchy's integral formula we can conclude that the considered covariant regularization of the spin-0 Feynman propagator admits the K\"all\'en-Lehman representation
\begin{equation}
   (2\pi)^4 i \propF(k) = \int_0^\infty {\rho(s)\over k^2 
       + s - i\epsilon}\, ds
   \label{KLrep}
\end{equation}
with
\begin{equation}
   \rho(s) = \delta(s-m^2) + (2\pi i)^{-1} \lim_{y \to 0} [
      \varphi(-s-iy) - \varphi(-s+iy) ] \,,
   \label{KLrho}
\end{equation}
$s, y > 0$. Note that $\rho(s)$ is real, $\rho(s) = O(s^{-r})$ as $s \to \infty$, and
\begin{equation}
   \int_0^\infty s^m \rho(s)\, ds = 0
   \qquad\hbox{for}\quad m = 0, 1,\ldots < r-1 \,.
   \label{KLcon}
\end{equation}
So we can decompose the regularized spin-0 propagator $\prop(x)$ into positive and negative energy parts: $\prop(x) = \Theta(x_0) \Delta^+(x) + \Theta(-x_0) \Delta^-(x)$\cite{Hooft}. We can analogously regularize each of the two terms of spin-$1/2$ and of spin-$1$ Feynman propagators so that they acquire no additional singularities and have the K\"all\'en-Lehman representation: in $x$-space we can decompose so regularized Feynman propagators into positive and negative energy parts without contact terms\cite{Hooft}.

Function $-i (2\pi)^{-4} ( \sqrt{\Lambda - m^2} + \Lambda)^n (k^2 + m^2 - i\epsilon)^{-1} (\sqrt{k^2 + \Lambda - i\epsilon} + \Lambda )^{-n}$, $\Lambda > m^2$, $n = 1, 2, \ldots $, is an example of a regularized spin-0 Feynman propagator that satisfies the above conditions with $r = n/2 + 1$. Unfortunately, we cannot use analogous propagators for a realistic regularization of QED since we do not know how to construct the corresponding \it local \rm Lagrangian.

In view of relations (\ref{KLrep}) and (\ref{KLrho}), we can regard a regulator that is an analytic function of the complex variable $k^2$ with a finite discontinuity across the negative real axis as a unitary regulator that is a limit of a weighted sum of propagators of the same kind as their number tends to infinity. Thus to use such regulators for a realistic regularization of QED, we have to generalize the method of unitary regulators\cite{Hooft} to an infinite number of auxilliary fields by constructing local, covariant Lagrangians whose propagators for interacting fields can be taken as regularized photon and electron propagators that have no additional singularities.

\section{Lagrangian for regularization of propagators in QED}

We will consider QED where photons have a very small mass $\mu$\cite{Hooft,Veltm}: With the source terms included, its Lagrangian reads
\begin{equation}
   \Lag_{QED} = -{\textstyle{1\over 4}} (\partial_\mu A_\nu 
	- \partial_\nu A_\mu)^2 - {\textstyle{1\over 2}} \mu^2 A^2 -
	 \ab{\psi} (\gamma^\mu \ldp_\mu + m )\psi + ie \ab{\psi} 
	 \gamma^\mu\psi A_\mu + A_\mu J_\mu + \ab{J_e}\psi 
         + \ab{\psi}J_e\,,
   \label{QEDL}
\end{equation}
where $J_\mu(x)$, $\ab{J_e}(x)$, and $J_e(x)$ are four-vector and bispinor source fields. The Feynman propagator for the four-vector field $A_\mu(x)$ is 
\begin{mathletters}
\label{PDeq}
\begin{equation}
   -i (2\pi)^{-4} { \delta_{\mu\nu} + \mu^{-2} k_\mu k_\nu \over
      k^2 + \mu^2 - i\epsilon } \,;
   \label{PDequ}
\end{equation}
and the Feynman propagator for the bispinor field $\psi(x)$ is
\begin{equation}
   -i (2\pi)^{-4} { -i \gamma^\mu k_\mu + m \over k^2 + m^2 
	- i\epsilon } \,.
   \label{PDeqv}
\end{equation}
\end{mathletters}

It is possible to modify the QED Lagrangian (\ref{QEDL}) so that the propagators for $A_\mu$ and $\psi$ fields are such regularizations of propagators (\ref{PDeq}) that have no additional singularities. Take, for example, the following real-valued, local Lagrangian\cite{mi001}:
\begin{mathletters}
\label{TTL}
\begin{equation}
   \LagT = - \Lag_1 - \Lag\DD + ie\ab{\psi} \gamma^\mu \psi A_\mu 
         + A_\mu J_\mu + \ab{J_e}\psi + \ab{\psi}J_e
   \label{TTLdf}
\end{equation}
with
\begin{eqnarray}
   \Lag_1 &\equiv& q_1^{-1} \int d^4p\, \Psi_\mu'(x,-p)
      [ \Lambda t(p^2) + p^\mu \ldp_\mu ] \Psi^\mu(x,p)
      \nonumber\\
   &&\qquad{}+ q_1^{-1} s_1 \int d^4p\, d^4p'\, f({p'}^2) f(p^2)
      [ \Psi_\mu'(x, -p')\Psi^{\prime\mu}(x,p) + p'_\nu p^\nu \Psi_\mu
        (x,-p') \Psi^\mu(x,p)
      \nonumber\\
   &&\qquad\quad{}- {p'}^\mu \Psi'_\mu(x,-p) p^\nu \Psi_\nu(x,p) ]
      \label{vecL}\\
   \Lag\DD &\equiv& q\DD^{-1} \int d^4p\, \ab{\Psi}\DD(x,-p)
      [ \Lambda t(p^2) + p^\mu\ldp_\mu ] \Psi\DD(x,p)
      \nonumber\\
      &&\qquad{}- q\DD^{-1} s\DD \int d^4p'\, d^4p\, f({p'}^2)
        f(p^2) [ \ab{\Psi}\DD(x,-p') \gamma^\mu \Psi\DD(x,p) p_\mu
         + {\rm c.c.} ] \,,
   \label{spinL}
\end{eqnarray}
\begin{equation}
   A_\mu(x) \equiv \int d^4p f(p^2) \Psi_\mu(x,p) \,,\qquad
   \psi(x) \equiv \int d^4p f(p^2) \Psi\DD(x,p) \,,
   \label{macvar}
\end{equation}
\end{mathletters}
where $\Psi_\mu(x,p)$ and $\Psi_\mu'(x,p)$ are four-vector-valued functions of two four-vectors $x$ and $p$; $\Psi\DD(x,p)$ is a bispinor-valued function of $x$ and $p$; $2 a \ldp_\mu b \equiv a (\partial_\mu b ) - (\partial_\mu a) b $; $\ab{\Psi}\DD \equiv \Psi\DD^\dagger \gamma^4$; $t(p^2)$ and $f(p^2)$ are real-valued functions of real $p^2$, $\int d^4 p f^2(p) = 1$; and $q_1$, $s_1$, $q\DD$, $s\DD$, and $\Lambda$ are real parameters. This Lagrangian is covariant and its own conjugate\cite{Hooft}.

Lagrangian $\LagT$ is a sort of Lagrangian defining unitary regulators\cite{Hooft} since it is constructed by modifying $\Lag_{QED}$ in the following sense: (i)~We introduced an infinite number of four-vector and bispinor fields of $x$ that have a continuous index $p$. (ii)~We replaced the free part of $\Lag_{QED}$ with the Lagrangian $-\Lag_1 - \Lag\DD$, which is of the first order in $\partial$ and has a nondiagonal mass matrix. (iii)~In the interaction and source terms, we replaced the fields $A(x)$ and $\psi(x)$ in $\Lag_{QED}$ with weighted integrals (\ref{macvar}) of $\Psi(x,p)$ and $\Psi\DD(x,p)$ over the continuous index $p$.

Consistent with (\ref{macvar}) we write
\begin{mathletters}
\label{varrep}
\begin{equation}
   \Psi_\mu(x,p) = \psi_\mu(x,p) + f(p^2) A_\mu(x) \,, \qquad
   \Psi\DD(x,p)  = \psi\DD(x,p) + f(p^2)\psi(x) \,,
\end{equation}
where $\psi_\mu(x,p)$ and $\psi\DD(x,p)$ are such that 
\begin{equation}
   \int d^4p f(p^2) \psi_\mu(x,p) = 0 \,,\qquad
   \int d^4p f(p^2) \psi\DD(x,p) = 0 \,, 
\end{equation}
\end{mathletters}
and consider $\psi_\mu(x,p)$, $A_\mu(x)$, $\Psi'(x,p)$, $\psi\DD(x,p)$, and $\psi(x)$ in $\LagT$ as independent fields. Proceeding as in Ref.~\onlinecite{mi001}, we can infer that the Feynman propagator for the four-vector field $A_\mu(x)$ in $\LagT$ equals
\begin{equation}
   -i (2\pi)^{-4} \Four{g}_1 \, {\delta_{\mu\nu} + 
      \Four{\mu}^{-2} k_\mu k_\nu \over k^2 + \Four{\mu}^2 }
   \label{phpro}
\end{equation}
with
\begin{equation}
   \Four{g}_1(k^2) \equiv q_1 s_1^{-2} I_{10} I_{20}^{-2}\,, \qquad
   \Four{\mu}(k^2) \equiv |s_1| I_{20} \,,
\end{equation}
\begin{equation}
   I_{mn}(k^2) \equiv 2 \pi^2 \Lambda^{-m} \int_0^\infty y^{m+n}
      f^2(y) t^{-m}(y) [ \sqrt{ 1 + \Lambda^{-2} k^2 y t^{-2}(y)}
      + 1]^{-m} dy \,;
\end{equation}
and the Feynman propagator for the bispinor field $\psi(x)$ in $\LagT$ equals
\begin{equation}
   -i (2\pi)^{-4} \Four{g}\DD \, { - i\gamma^\mu k_\mu + \Four{m} \over
       k^2  + \Four{m}^2 }
   \label{elpro}
\end{equation}
with
\begin{equation}
   \Four{g}\DD(k^2) \equiv q\DD s\DD^{-1} I_{10}I_{20}^{-1}\,, 
      \qquad \Four{m}(k^2) \equiv s\DD^{-1} \{ 1 - s\DD^2 [
      I_{10} I_{11}  + {\textstyle{1\over4}} k^2 I_{20}^2 ] \}
      I_{20}^{-1} \,;
\end{equation}
where $k^2$ has to be replaced everywhere with $k^2 - i\epsilon$, by the Feynman prescription\cite{Hooft}.

If functions $t(p^2)$ and $f(p^2)$ are such that
\begin{equation}
   \int_0^\infty f^2(y) t(y) | \sqrt y /t(y)|^{l+1} dy = 0
   \label{ftpog}
\end{equation}
for $l=0, -1, \ldots,-n$, then for complex values of $k$ as $|k^2| \to \infty$:
\begin{equation}
   \left| \Four{g}_1 \, {\eta_{\mu\nu} - \Four{\mu}^{-2} k_\mu k_\nu
      \over \Four{\mu}^2 + k^2 } \right| = O(|k^2|^{(1-n)/2})
   \label{elreg}
\end{equation}
and
\begin{equation}
   \left| \Four{g}\DD \, {\Four{m} - i\gamma^\mu k_\mu 
      \over \Four{m}^2 + k^2 } \right| = O(|k^2|^{-n/2}) \,.
   \label{phreg}
\end{equation}

For any $\mu^2$, $m^2$ and integer $n$, there are real functions $f(p^2)$ and $t(p^2)$, and constants $s_1$, $s\DD$, $q_1$, $q\DD$, and $\Lambda_0 > 0$ such that the propagators (\ref{phpro}) and (\ref{elpro}) with $\Lambda > \Lambda_0$ are regularizations of spin-1 and spin-$1\over 2$ propagators (\ref{PDeq}) analogous to (\ref{scpro})\cite{foot2}. In such a case: (i)~$I_{mn}(k^2)$ are real for $k^2 \ge -m^2$, (ii)~The constants $s_1$, $s\DD$, $q_1$ and $q\DD$ are such that
\begin{equation}
   \Four{\mu}^2(-\mu^2) = \mu^2 \,, \qquad
   \Four{m}^2(-m^2) = m^2
   \label{cond1}
\end{equation}
\begin{equation}
   \Four{g}\DD(-\mu^2) = 1 + d\Four{\mu}^2(k^2 = -\mu^2)/ dk^2 \,,
   \qquad
   \Four{g}_1(-m^2) = 1 + d\Four{m}^2(k^2 = -m^2)/ dk^2 \,.
   \label{cond2}
\end{equation}
So propagators (\ref{phpro}) and (\ref{elpro}) with $\epsilon = 0$ have poles at $k^2 = -\mu^2 $ and $k^2 = - m^2$, respectively, and their behaviour at these poles is given by spin-1 and spin-$1\over 2$ propagators (\ref{PDeq}) with $\epsilon = 0$. (iii)~The difference between spin-1 propagator and propagator (\ref{phpro}) can be made arbitrarily small for any $|k^2| \le K$ for all $K > 0$, if we choose $\Lambda$ sufficiently large; and the same goes for spin-$1\over 2$ and (\ref{elpro}) propagators. So we may consider $\Lambda $ as a cut-off constant. (iv)~The propagators (\ref{phpro}) and (\ref{elpro}) are analytic functions of $k^2$ that (a)~are not continuous across the negative real axis, (b)~have no additional singularities to those of spin-1 and spin-$1\over 2$ propagators (\ref{PDeq}), and (c)~satisfy relations (\ref{elreg}) and (\ref{phreg}).

\section{An extension of QED that does not need to be regularized}

The Lagrangian $\LagT$ as defined by (\ref{TTL}) and (\ref{varrep}) has the same interaction and source terms as the QED Lagrangian $\Lag_{QED}$. So we define the quantum field theory based on $\LagT$ (trQED) by the n-point Green functions and corresponding $S$-matrix elements expressed in terms of QED diagrams where the spin-1 and spin$1\over 2$ propagators (\ref{PDeq}) are replaced with such propagators (\ref{phpro}) and (\ref{elpro}) that are their regularization. This definition of trQED is consistent with the 't Hooft and Veltman definition of a quantum field theory directly in terms of Feynman diagrams corresponding to its Lagrangian\cite{Bjork}.

trQED is a realistic quantum field theory, because (i)~trQED propagators (\ref{phpro}) and (\ref{elpro}) have no additional singularities and admit the K\"all\'en-Lehman representation, and (ii)~trQED n-point Green's functions are causal and the corresponding $S$-matrix is unitary to any order in the fine structure constant. To show this we may follow the 't Hooft and Veltman proof for the case of unitary regulators\cite{Hooft}.

In the asymptote $\Lambda \to \infty$, the propagators (\ref{phpro}) and (\ref{elpro}) of trQED behave by (\ref{regpr0}) and (\ref{regpr}) the same way as unitary regularized spin-1 and spin$1\over 2$ propagators (\ref{PDeq}) when the auxilliary masses tend to infinity. So we can compute the renormalized n-point Green's functions of QED from the n-point Green functions of trQED by renormalization, choosing appropriate dependence of $e$, $s_1$, $s\DD$, $q_1$, and $q\DD$ on $\Lambda$, and then limiting $\Lambda \to \infty$. So trQED is at least as good as QED for modeling quantum electrodynamic phenomena: \it trQED is a realistic extension of QED that requires no regularization, \rm while QED can be regarded as a physically appropriate simplification of the more complicated trQED by renormalization.

The question remains what physical phenomena can we model by the trQED Lagrangian better than by the QED Lagrangian. E.g.: (i)~Up to any value of $|k^2|$ the differences between photon and electron propagators (\ref{PDeq}) and their regularizations (\ref{phpro}) and (\ref{elpro}) can be made as small as desired if the cut-off parameter $\Lambda$ is sufficiently large. In such a case, the propagators (\ref{PDeq}) provide a low-energy approximation to their regularizations, the trQED propagators (\ref{phpro}) and (\ref{elpro}) whose ultra-high-energy behaviour looks physically more appropriate---we do not need to alter (regularize) it to make the definition of trQED formally correct. As trQED propagators together with vertices determine trQED cross sections, we may expect trQED to better model quantum-electrodynamic phenomena at higher energies than QED. The question is how much better, especially if we use dressed propagators instead of the bare ones, (\ref{phpro}) and (\ref{elpro}), in computing the diagrams of trQED. (ii)~Ever since the EPR gedanken-experiment, it is known that interpretations of certain quantum phenomena suggest existence of faster-than-light effects. The question is whether we can model them using the Euler-Lagrange equations of $\LagT$ whose retarded solutions have unbounded front velocities, see Ref.~\onlinecite{mi001}. 

\section{acknowledgement}

Authors greatly appreciate discussions with M. Polj\v sak and his suggestions.


\begin{references}
\bibitem{foot2}To show this we considered the case where function $t(z)/\sqrt z$ takes only a finite number of real values of approximately the same magnitude.
\bibitem{foot5}The fact that there are functions $f(p^2)$ and $t(p^2)$ and constants $s_1$, $s\DD$, $q_1$, $q\DD$, $\Lambda$, and $e$ such that trQED is as good as QED for modeling electrodynamic quantum phenomena in general and probably better for modeling ultra-high-energy ones gives support to our hypothesis\cite{mi001}: The Euler-Lagrange equations of a theory that underlies the standard model in the four-dimensional spacetime formally resemble some equations of motion underlying fluid-dynamic equations in the kinetic theory of gases.
\end{references}
\end{document}